\documentclass{easychair}

\usepackage{doc}

\title{Information Credibility in the Social Web:\\Contexts, Approaches, and Open Issues\footnote{Article accepted and presented at ITASEC 2020: Italian Conference on Cybersecurity. February 4-7, 2020, Ancona, Italy. \url{https://itasec.it/}}}

\author{
Gabriella Pasi 
\and
    Marco Viviani
}

\institute{
  University of Milano-Bicocca\\Department of Informatics, Systems, and Communication\\Information and Knowledge Representation, Retrieval, and Reasoning LAB\\ \url{http://www.ir.disco.unimib.it}\\
  \email{\{gabriella.pasi, marco.viviani\}@unimib.it}
 }

\authorrunning{Pasi and Viviani}

\titlerunning{Information Credibility in the Social Web}

\begin{document}

\maketitle

\begin{abstract}
In the Social Web scenario, large amounts of User-Generated Content (UGC) are diffused through social media often without almost any form of traditional trusted intermediaries. Therefore, the risk of running into misinformation is not negligible. For this reason, assessing and mining the credibility of online information constitutes nowadays a fundamental research issue. Credibility, also referred as believability, is a quality perceived by individuals, who are not always able to discern, with their own cognitive capacities, genuine information from fake one. Hence, in the last years, several approaches have been proposed to automatically assess credibility in social media. Many of them are based on data-driven models, i.e., they employ machine learning techniques to identify misinformation, but recently also model-driven approaches are emerging, as well as graph-based approaches focusing on credibility propagation, and knowledge-based ones exploiting Semantic Web technologies. Three of the main contexts in which the assessment of information credibility has been investigated concern: $(i)$ the detection of opinion spam in review sites, $(ii)$ the detection of fake news in microblogging, and $(iii)$ the credibility assessment of online health-related information. 
In this article, the main issues connected to the evaluation of information credibility in the Social Web, which are shared by the above-mentioned contexts, are discussed. A concise survey of the approaches and methodologies that have been proposed in recent years to address these issues is also presented.
\end{abstract}



%
%

\section{Introduction}
\label{sect:introduction}

In the `offline' world, 
long before the Web became publicly available, users could rely on traditional forms of information verification, such as the presence of traditional media \emph{intermediaries} such as experts, by considering their \emph{reputation}, or \emph{trust} them based on first-hand experiences. Nowadays, in the Social Web scenario, almost everyone can spread contents on social media in the form of \emph{User-Generated Content} (UGG), almost without any traditional form of trusted control \cite{ferrari2013privacy}. This `disintermediation' process \cite{Eysenbach2008} has lead, on the one hand, to the democratization of the information diffusion, but, on the other hand, to the spread of possible fake news and misinformation, which risk to severely affect social aspects of living.

We live, in fact, in a so-called `post-truth' era, in which objective facts are less influential in shaping public opinion than appeals to emotion and personal belief. This is partially due to the fact that, in social media, user-created networks can become real \emph{echo chambers} \cite{jamieson2008echo}, in which one point of view dominates all the others, the verification of the statements has usually no effect, and this allows the repetition of unverified statements without refutation. This is connected to the cognitive phenomenon of \emph{confirmation bias} that, in psychology, is defined as the tendency to search for, interpret, favor, and recall information in a way that affirms one's prior beliefs or hypotheses \cite{plous1993psychology}. Furthermore, the echo chamber phenomenon is emphasized by the filtering algorithms that are the basis of social media in proposing information of interest: by suggesting personalized (information) items that consider different elements of the user profile, such as location, past click-behavior and search history, users become separated from information that disagrees with their viewpoints, effectively isolating them in their own cultural or ideological bubbles, the so-called \emph{filter bubbles} \cite{pariser2011filter}. The echo chamber and filter bubble phenomena are currently being studied by several researchers, in different fields \cite{burbach2019bubble,del2015echo,sasahara2019inevitability}.

In this scenario, it becomes essential to try to find and develop automatic solutions (i.e., algorithms, user interfaces, search engines, mobile apps) that assist the user to get out of their echo chambers and become aware of the level of \emph{credibility} of the information they come into contact with. In particular, the contexts in which the problem of credibility assessment is receiving the most interest are those of \emph{opinion spam} detection, \emph{fake news} detection and evaluation of \emph{health-related} information that spreads online. 

Opinion spam refers to malicious activities  attempting to mislead  either  human readers, or automated  opinion mining and sentiment  analysis systems. In most cases, it concerns \emph{fake reviews}, whose spread has negative effects on both businesses and potential customers \cite{jindal2008opinion}. Fake news are particularly diffused in microblogging platforms, where  millions of users act as  real-time news diffusers, since the so-called \emph{trending topics} can be considered, in all respects, ``headline news or persistent news'' \cite{kwak2010twitter}. In this context, detecting fake news is particularly relevant to prevent public opinion from being manipulated \cite{conroy2015automatic,de2019fake}. Concerning health-related information, numerous are the communities of individuals who share contents about personal health problems or try to carry out self-diagnosis through Question-Answering (QA) systems or professional healthcare services allowing people to interact online \cite{hajli2014developing}. In this case, incurring in inaccurate, unverified, unsuitable information can lead to potentially very harmful consequences for individuals with a low health literacy \cite{bates2006effect}.

In all three of these contexts, multiple solutions have been studied or are being studied to solve context-specific problems related to the credibility of the information diffused. However, the approaches proposed so far in the literature share common characteristics that allow to classify them from a more general point of view, as illustrated in this article, which is organized as follows. Initially, the concept of credibility will be presented, addressed both from the point of view of the social/communication sciences and within the Social Web scenario (Section \ref{sec:offlineonline}); then, a  classification of the main state-of-the-art approaches for the automatic or semi-automatic evaluation of the credibility of online information will be provided (Section \ref{sec:approaches}); finally, the open issues that remain to be faced for the development of effective solutions to the considered problem will be highlighted, as well as some further research directions (Section \ref{sec:openissues}).

\section{From `Offline' to `Online' Credibility Assessment}\label{sec:offlineonline}

The study of the credibility of information has a long history and concerns different contexts and research fields, such as social sciences (including in particular communication and psychology), and computer science, also at interdisciplinary level. Over the years, the interest has gradually moved from traditional communication environments, characterized by interpersonal and persuasive communication, to mass communication and interactive-mediated  communication, with particular  reference to online communication \cite{metzger2007making}. 

Therefore, in this section, the main aspects related to the evolution of the study of information credibility are briefly discussed, focusing, in particular, on two different scenarios: $(i)$ the `offline' scenario, where information is disseminated through means of `traditional' media that do not require the use of digital technologies, and $(ii)$ the `online' scenario, where information is diffused through the (Social) Web by means of Web 2.0 technologies.

\subsection{Offline Information Credibility}
In the research field of communication, the notion of \emph{credibility} has been investigated since ancient times. In fact, among the first works that have come down to us that discuss this concept, there are the \emph{Phaedrus} by Plato, and the Aristotle's \emph{Rhetoric}, both dating back to the 4th Century BC, where the focus is, in particular, on the study of the characteristics of the speaker that make her/him believe credible. Over the years, depending on the context, credibility has been associated with other many related concepts or combinations of them, such as believability, trustworthiness, perceived reliability, expertise, accuracy, etc. An in-depth analysis of the nuances of meaning related to these concepts is provided in \cite{self2014credibility}. 

In the modern era, in the fields of psychology and communication, the research undertaken in the 1950s by Hovland \emph{et al.} \cite{hovland1953communication} constitutes the first systematic work about credibility and mass media, focusing specifically on the study of the impact that the credibility of the information source, i.e., \emph{source credibility}, has on the assessment of the credibility of  mass communication messages. The authors investigated, in particular, how people evaluated these messages coming from ``high credibility'' and ``low credibility'' sources. 

Later on, Fogg and Tseng in \cite{fogg1999elements} stated that credibility, as a \emph{perceived} quality of the \emph{information receiver}, it is composed of \emph{multiple dimensions} that have to be considered. In this sense, the process of assessing information credibility involves different \emph{characteristics}, which can be connected to: $(i)$ the \emph{source} of information, $(ii)$ the information that is diffused, i.e., the \emph{message}, considering both its structure and its content, and $(iii)$ the \emph{media} used to diffuse information \cite{self2014credibility}.

\subsection{Online Information Credibility}\label{sec:credibilitysocialweb}

Online, and in the Social Web in particular, information credibility assessment deals with the analysis of both UGC and their authors' characteristics \cite{Moens14}, and the intrinsic community-based nature of social media platforms \cite{Safko:2010:SMB:1941342}. Specifically, this means to take into account \emph{credibility features} connected to: $(i)$ the \emph{users} forming a virtual community (i.e., the information sources), $(ii)$ their \emph{User-Generated Content} (i.e., the information the users diffuse), and $(iii)$ the \emph{social relationships} connecting the users in the virtual community (i.e., the main characteristic of an online social networking system), also considering other kinds of entities possibly connected to users with other kinds of relationships \cite{carminati2012multi}. 

The problem of evaluating the above-mentioned aspects, in a scenario characterized by a large amount of information generated directly by users such as the Social Web, becomes particularly complex for information receivers. While in previous media contexts people usually dealt with traditional trusted intermediaries on which they could rely on, e.g., reputed newspapers, popular enterprises, no profit organizations, etc., social media ``have in many ways shifted the burden of information evaluation from professional gatekeepers to individual information consumer'' \cite{FlanaginMet2008}. This constitutes a tricky problem because, even if credibility is a characteristic perceived by individuals, credibility assessments can be very complex for human beings, especially in the online environment, because the effort it takes to verify the credibility of Web-based information would be disproportionate and often ineffective \cite{metzger2007making}.
In fact, humans have limited cognitive capacities to effectively evaluate the information they receive, especially in situations where the complexity of the features involved and of their interactions increases \cite{lang2000limited}. 

For the above-mentioned reasons, there is nowadays the need of developing interfaces, tools or systems that are designed to help users in automatically or semi-automatically assess information credibility. 
In the next section, an high-level classification of the approaches that have been proposed so far to tackle this issue in the Social Web is provided.

\section{Approaches to Credibility Assessment}\label{sec:approaches}

A first classification that can be made with respect to recent approaches that assess the credibility of online information concerns the implementation of \emph{data-driven} or \emph{model-driven} approaches. In the first case, starting from some available data, a bottom-up model is learned to identify credible information with respect to non-credible one. In the second case, some domain knowledge is available, which is used to build a top-down model to tackle the considered issue.
Another classification dimension that can be considered, concerns the fact of studying the \emph{propagation} of (false) online information or the attempt to produce a \emph{classification} (or a \emph{ranking}) of information based on its credibility level.

As illustrated in Section \ref{sect:introduction}, the approaches that fall into the above-mentioned categories are used in different research contexts to the evaluation of the credibility of online information, such as opinion spam detection, fake news detection, and credibility assessment of online health-related information \cite{viviani2017credibility}. Although each of these contexts has its own peculiarities, the common aspects that remain valid for each category of approaches will be explained in the following.

\subsection{Information Propagation} In general, \emph{propagation-based approaches} are mainly concerned with studying the influence that \emph{spammers} or \emph{social bots} have on the dissemination of (false) information or how low-credibility information spreads over the social network structure \cite{shao2018spread}. The identification of genuine (or, conversely, non-credible) information deals, in this case, mostly with the identification of these malicious users, or with the detection of suspicious behaviors in the information diffusion. 

These approaches usually rely on the analysis of the graph-based representation of the social network structure, and often employ unsupervised learning algorithms to detect cliques of malicious users (e.g., via Social Network Analysis techniques and metrics) and/or to identify the so-called \emph{bursts}, i.e., sudden increases of the use of particular sets of keywords in a very short period of time \cite{fei2013exploiting}. Propagation-based approaches can also rely on some pre-computed credibility values (usually learned from a classifier) and study their spread over the social network structure. 

The study of malicious users' behaviors and false information propagation has in many cases a slightly different purpose (even if it is closely related) with respect to the assessment of information credibility (meant as the valuation of information items), which is more related to classification-based approaches, which are illustrated below.

\subsection{Information Classification} 

In \emph{classification-based approaches}, fall both data-driven methods that are based on the use of (mostly supervised or semi-supervised) learning algorithms, and model-driven approaches which are based on some prior \emph{domain knowledge}; both of them aim at classifying in a binary way information items (i.e., credible VS non-credible) or at producing a ranking based on an estimated credibility probability or a credibility score computed for each item. The model-driven category includes both the use of the Multi-Criteria Decision Making (MCDM) paradigm \cite{greco2016multiple}, and the use of Knowledge Bases and Semantic Web technologies \cite{giaretta1995ontologies}.

\subsubsection{Approaches based on Credibility Features}
When considering machine learning techniques and the MCDM paradigm to assess information credibility, different characteristics, namely \emph{credibility features}, are taken into account. As illustrated in Section \ref{sec:credibilitysocialweb}, these features are generally related to the users in the virtual community, the information items that are generated and diffused, and the virtual relationships among users and other entities in the community. For this reason, it is possible to provide the following classification:
\begin{itemize}
    \item \emph{Behavioral features}: they are related to the users generating and diffusing information. They can be extracted both from public Web data, e.g., user ID, usual time of posting, frequency of posting, presence of a public image profile, etc., and private/internal Web data, e.g., IP and MAC addresses, time taking to post an information item, physical location of the user, etc. \cite{mukherjee2013spotting}.
    \item \emph{Content-based features}: they are related to the textual content of the information item. They can be both lexical features such as word $n$-grams, part-of-speech, and other lexical attributes, and stylistic features, e.g., capturing content similarity, semantic inconsistency, use of punctuation marks, etc. \cite{kdir19}
    \item \emph{Social (Graph-based)} features: they capture complex relationships among users, the information they diffuse, and other possible entities (e.g., products and stores) in the social network \cite{Shu:2017:FND:3137597.3137600}.
\end{itemize}

\medskip\noindent{\bf Data-driven Approaches.} In the case of using data-driven approaches that employ well-known (supervised) machine learning techniques (e.g., SVM, Random Forests, etc.), often a binary classification of information items with respect to their credibility is obtained by training a model over the considered set of features and one or more suitable labelled dataset(s). These approaches are hence based on a feature extraction and selection phase, and are dependent on the availability of (unbiased) labelled data, which is not always the case, as illustrated in the literature \cite{viviani2017credibility}.
Furthermore, some of the machine learning techniques proving to be effective in the considered research field, are often inscrutable to observers (they are characterized by the so-called `black-box' behavior), making it difficult to evaluate the importance of distinct and/or interacting features in obtaining the classification results. A possible solution would be the study of approaches based on the so-called \emph{eXplainable Artificial Intelligence} (XAI), referring to methods and techniques in the application of AI technology such that the results of the solution can be understood by human experts \cite{holzinger2018machine}.

\medskip\medskip\noindent{\bf Model-driven Approaches.} Approaches for assessing credibility that allow the human being to be involved in the decision process and to (possibly) understand the result obtained, are those modeling the considered problem as a Multi-Criteria Decision Making problem, characterized by the presence of a set of \emph{alternatives}, i.e., the information items to be evaluated in terms of credibility, and a set of \emph{criteria}, i.e., the considered set of credibility features. In an MCDM problem, each alternative `satisfies' each criterion to a certain extent, producing this way a performance score, i.e., the credibility score, one for each criterion associated with the alternative.
For each alternative, the \emph{aggregation} of these multiple credibility scores produces an overall performance score, i.e., an overall credibility score. 

In such kind of modeling of the problem, there is usually a prior knowledge of the features to be considered, and of the \emph{importance} that each feature has in terms of credibility (e.g., this information can be taken from prior literature specifically addressing the problem from a social and economic point of view \cite{luca2016fake}). Furthermore, some \emph{satisfaction functions} are defined to transform the values of the features into suitable credibility scores. Finally, suitable \emph{aggregation functions} to obtain the overall credibility scores can be selected or defined. This way, having an overall credibility score associated with each information item, it is possible to provide both: 
\begin{itemize}
    \item A \emph{binary classification} of information items into genuine/fake (by selecting a suitable classification threshold, as illustrated in \cite{sebastiani2005text}).
    \item A \emph{ranking} of the information items based on their overall credibility score.
\end{itemize}

Different an numerous are the families of aggregation functions to be considered for tackling the problem, depending on the preferences of the decision maker \cite{viviani2016multi,viviani2017quantifier}. \emph{Ordered Weighted Averaging} (OWA) operators, in particular, have been extensively studied in the literature \cite{yager2012ordered}. In fact, they give the possibility to guide the aggregation by \emph{linguistic quantifiers} (e.g., \emph{all}, \emph{some}, \emph{many}, ...), which allow the decision maker to obtain the best alternative(s) based on the satisfaction of a ‘certain amount’ of the criteria by the alternative(s). Recently, an MCDM approach has been proposed that allows to model \emph{interacting} features, by employing the \emph{Choquet integral} \cite{pasi2019multi}; in this work, and in general in MCDM approaches based on aggregation operators, it can be complex to define the model when the number of features increases \cite{grabisch2010decade}.

\subsubsection{Approaches based on Knowledge Bases} Another way to use prior domain knowledge to assess the credibility of online information is to refer to the use of Knowledge Bases \cite{shi2016fact}. In this context, knowledge is not referred to what are the credibility features associated with information, and their importance. In this case, the information that is known to be credible is available, and it is expressed in terms of \emph{facts}.
This type of approach is used in particular for \emph{automated fact checking} \cite{thorne2018automated}, in situations where: $(i)$ manual information credibility assessment is not feasible, e.g., the number of human experts is too limited with respect to the amount of information to be verified; $(ii)$ crowdsourcing-based information credibility assessment does not guarantee the credibility of assessors.

Technically speaking, a \emph{fact} is a set of (Subject, Predicate, Object) (SPO) triples extracted from texts that well-represent a given information. E.g., ``Giacomo Puccini is a composer'' can be expressed as (GiacomoPuccini, Profession, Composer). Facts must be processed and cleaned up (i.e., redundant, outdated, conflicting, unreliable or incomplete data are removed) to build a \emph{Knowledge Base}, i.e., a set of SPO triples. A graph structure, known as the \emph{knowledge graph}, can be used to represent the SPO triples in a Knowledge Base, where the entities (i.e., subjects or objects in SPO triples) are represented as nodes, and relationships (i.e., predicates in SPO triples) are represented as edges. Knowledge Bases are suitable candidates for providing \emph{ground truth} to information credibility assessment studies, i.e., it is possible to reasonably assume the existing triples in a Knowledge Base represent true facts.

Making these premises, to evaluate the credibility of to-be-verified information items, in turn represented as SPO triples, it is sufficient to compare them with the SPO triples contained in the Knowledge Base(s), obtaining, in this way, a classification of the information items. Open issues connected to this kind of approach concern how to consider missing facts in the Knowledge Base(s), and, connected to this problem, how to constantly update the Knowledge Base(s) with up-to-date credible information.

\section{Open Issues and Further Research}\label{sec:openissues}

As illustrated in this paper, numerous approaches belonging to different categories have been proposed up to now to assess information credibility in the Social Web, with respect to distinct research contexts, i.e., opinion spam detection, fake news detection, health-related information credibility assessment. Every category of approaches presents both advantages and drawbacks, which are summarized in the following.

\emph{Propagation-based approaches} allow to effectively identify misinformation spammers and spam bots, leading to the possible identification of false information, but they are affected by the problem of managing and analyzing complex structures such as graphs, and, in some cases, by the need of having pre-computed credibility values to capture and investigate how (genuine or fake) information spreads over the network.

\emph{Classification-based approaches} are affected by the possible inscrutability of results and data dependency in the case of supervised `black-box' approaches, and by the difficulty of managing the complexity of the model when the number of criteria increases in the case of the MCDM scenario; despite this, both turned out to be particularly effective in well defined tasks \cite{viviani2017credibility}. 

In approaches based on the use of \emph{Knowledge Bases}, some issues emerge about the treatment of missing information, conflict resolution and triples update in the knowledge graph(s), especially in the Social Web scenario where new information is produced with a very high frequency and volume. Despite this, they can be particularly useful and of particular support to automated fact checking. 

From a technical point of view, further research in the information credibility assessment field must deal with the above-mentioned issues, by developing, for example, \emph{hybrid approaches} able to simultaneously exploit domain knowledge, model-driven aspects, and supervised learning when unbiased training data are available, even of a small number. Moreover, a problem of particular relevance in recent years is that of \emph{early identification} of false information \cite{liu2018early} or of \emph{hate speech} \cite{davidson2017automated} (whose detection is a problem that is related to information credibility assessment), to act immediately to avoid the diffusion of contents that are harmful in several respects (e.g., racist, sexist, homophobic, scientifically inaccurate contents) before they can cause further damage to the community. 

Another problem that should be addressed in the near future is how to consider, in the credibility assessment process, the amount of \emph{confidential information} that is released within UGC, an aspect that has not been sufficiently investigated in the literature until now \cite{damiani2009trading,livraga2019data}. On the one hand, it is necessary to develop solutions to protect this type of information, on the other hand, how to use it (once properly sanitized) to have additional evidence about the reliability of the source of information and therefore of the credibility of her/his content.

From the point of view of the research contexts in which the problem has been addressed in recent years, it must be emphasized that, while the study of techniques for opinion spam detection and fake news detection has seen the scientific community particularly prolific \cite{heydari2015detection,sharma2019combating}, the study of the problem of evaluating the credibility of health-related information can still be considered at a preliminary level. In fact, at the time of writing, despite the great interest that this area has, there are still few solutions that started to evaluate the credibility of the health-related UGC spreading on social media, e.g., \cite{mukherjee2014people}. Most approaches have focused on assessing the credibility of the means of communication that disseminates the information, e.g., the credibility of websites or blogs that publish medical information \cite{bates2006effect,wang2008health}. Other works have done meta-analytical studies, e.g., \cite{ma2017user}. It is therefore necessary to develop automatic or semi-automatic approaches that help people not to incur the possible negative consequences that the social ``word-of-mouth" can have, especially in such a delicate context as that of health.

\label{sect:bib}
\bibliographystyle{plain}
\bibliography{easychair}
\end{document}